\documentclass{article}
\usepackage{latexsym}
\usepackage{amssymb}
\usepackage{amsmath}
\usepackage{graphicx}

\begin{document}

\title{On the relation of Thomas rotation and angular velocity of reference
frames}
\author{T.~Matolcsi\thanks
{Department of Applied Analysis, E\"otv\"os University, P\'azm\'any P.
s\'et\'any 1C., H--1117 Budapest, Hungary. Supported by OTKA-T 048489.},~
M.~Matolcsi\thanks
{Alfr\'ed R\'enyi Institute of Mathematics, Hungarian Academy of Sciences,
Re\'altanoda utca 13--15., H--1053, Budapest, Hungary. Supported by OTKA-PF
64061, T 049301, T 047276.} ~and~
T.~Tasn\'adi\thanks
{Department of Solid State Physics, E\"otv\"os University, P\'azm\'any P.
s\'et\'any 1A., H--1117 Budapest, Hungary. Supported by OTKA-F 43749.}}
\maketitle

\newcommand\uu{\mathbf u} \newcommand\g{\mathbf g}
\newcommand\Hh{\mathbf H} \newcommand\D{\mathrm D}
\newcommand\Eu{\mathbf E_\uu} \newcommand\N{\mathbf N}
\newcommand\h{\mathbf h} \newcommand\1{\boldsymbol 1}
\newcommand\x{\mathbf x}
\newcommand\y{\mathbf y}
\newcommand\I{\mathbf I} \newcommand\R{\mathbf R}
\newcommand\vv{\mathbf v}
\newcommand\M{\mathbf M}
\newcommand\V{\mathbf V}
\renewcommand\t{\mathbf t}
\newcommand\Pp{\mathbf P}
\newcommand\s{\mathbf s}
\newcommand\si{\sigma}
\newcommand\be{\begin{equation}}
\newcommand\U{\mathbf U}
\newcommand\q{\mathbf q}
\newcommand\Om{\boldsymbol\Omega}
\newcommand\E{\mathbf E}
\newcommand\z{\mathbf z} \renewcommand\L{\mathbf L}
\newcommand\zz{\hat{\mathbf z}}
\newcommand\A{\mathbf A}
\newcommand\uc{\uu_c}
\newcommand\as{\mathbf a}
\newcommand\al{\alpha}
\newcommand\hal{\hat\alpha}
\newcommand\bb{\beta}
\newcommand\hbb{\hat\beta}
\newcommand\dd{\mathbf d}
\newcommand\kk{\mathbf k} \newcommand\hh{\mathbf h}
\newcommand\RR{\mathbb R}
\newcommand\F{\mathbf F}
\newcommand\De{\Delta}

\begin{abstract}

In the extensive literature dealing with the relativistic phenomenon of
Thomas rotation several methods have been developed for calculating the
Thomas rotation angle of a gyroscope along a circular world line. One of
the most appealing concepts, introduced in \cite{rindler}, is to
consider a rotating reference frame co-moving with the gyroscope, and relate
the precession of the gyroscope to the angular velocity of the reference frame.
A recent paper \cite{herrera}, however, applies this principle to three
different co-moving rotating reference frames and arrives at three different
Thomas rotation angles. The reason for this apparent paradox is that the
principle of \cite{rindler} is used for a situation to which it does not apply.
In this paper we rigorously examine the theoretical background and limitations
of applicability of the principle of \cite{rindler}. Along the way we also
establish some general properties of {\it rotating reference frames},
which may be of independent interest.

\end{abstract}


\section{Introduction}

The relativistic phenomena of {\it Thomas rotation}  and
{\it Thomas precession} have been
treated in relativity theory from various points of
view (see e.g. \cite{costella},  \cite{fisher}, \cite{kennedy}, \cite{matolcsi},
\cite{moller}, \cite{philpott}, \cite{rebilas}, \cite{rindler}, \cite{thomas},
\cite{ungar},  \cite{wilkins}). (Unfortunately, there seems to be no
standard agreement in the literature as to the use of the terminologies of
rotation and precession; we will adhere to
the terminologies used in \cite{matolcsi}.) We remark that
these notions also provide a possible way to put relativity to the test in
practice: Thomas rotation is one of the relativistic effects currently being measured
in the Gravity Probe B experiment.

A brief overview of the appearing notions is as follows:

Consider Special Relativity first. As explained in \cite{matolcsi} in
detail, one must make a clear distinction between the notions of precession and
rotation. On the one hand, {\it Thomas precession}
refers to the instantaneous angular velocity, {\it with respect to a
particular inertial frame}, of a gyroscope moving along an arbitrary  world line. At every frame-instant the inertial frame Lorentz boosts the gyroscopic vector to its own space and sees the vector precess.

Thus, Thomas precession

-- \ makes sense for arbitrary gyroscopes with respect to arbitrary inertial
frames,

-- \ is a continuous phenomenon (i.e. it makes sense at every proper time instant
of the gyroscope),

-- \ is a {\it relative} notion, i.e. the same gyroscope may show different
instantaneous precessions with respect to different inertial frames.

On the other hand, {\it Thomas rotation} refers to the spatial rotation
experienced by a gyroscope which has {\it returned to its initial} local rest
frame.

Thus, Thomas rotation

-- \ makes sense only for `returning' gyroscopes (i.e. for proper time
instances $s$ such that the world line $r$ along which the gyroscope moves
satisfies  $\dot r(s)=\dot r(0)$),

-- \ is a discrete phenomenon (i.e. it makes sense only for the (usually)
discrete set of proper time instances when the gyroscope happens to be in
its initial local rest frame),

-- \ is an {\it absolute} notion, i.e. independent of who observes it
(the same angle of Thomas rotation will be measured by all inertial frames
observing the gyroscope).

In terms of any particular inertial frame one can think of Thomas rotation
as the time-integral of instantaneous Thomas precessions (and while Thomas
precession, as a function of time, will differ from one inertial frame to
another, its integral will always give the same angle: the Thomas rotation).

Consider now General Relativity.
Clearly, Thomas precession makes no sense here, because it requires inertial
frames (and Lorentz boosts).
On the other hand, Thomas rotation {\it can} make sense. (Let us note here
that in different spacetime models different names are given to the same
phenomenon, e.g. Fokker-de Sitter precession in Schwarzschild space,
Schiff precession in Kerr space; we find it reasonable to use the unified
terminology `Thomas rotation' throughout this paper.)
Since, in general, the tangent spaces at $r(s)$ and at
$r(0)$ are different, Thomas rotation can be meaningful {\it only if} the
tangent spaces in question are identified somehow, and
$\dot r(s)=\dot r(0)$ under this identification. (One also expects the
identification to be `natural' from a physical point of view; mathematically it
is not important. In special relativity this `natural' identification is simply
the identity map, as the tangent spaces are the same.) Such identification is
possible e.g. by the Lie transport corresponding to a Killing vector.

Such a situation is considered in \cite{rindler} where rotating coordinates
are suggested for calculating the Thomas rotation angle on a circular orbit.
The principle in that paper is that, heuristically, the gyroscope keeps
direction in itself, therefore if a rotating reference frame co-moving
with the gyroscope has instantaneous angular velocity $\Om$, then it will see
the gyroscope precess with angular velocity $-\Om$. Then, when the gyroscope
returns to its initial local rest frame one can evaluate the Thomas rotation
angle from the knowledge of instantaneous precessions along the way. Note that
this method is {\it conceptually} different from the special relativistic
evaluation of the rotation angle from knowledge of instantaneous Thomas precessions.
For a clear distinction:

-- \  Thomas precession involves a gyroscope, an inertial frame,
and the base-point of the gyroscope is moving in the space of the inertial
frame (and it makes sense only in special relativity).

-- \  The method suggested by \cite{rindler} involves a gyroscope,
a co-moving rotating frame such that the base-point of the gyroscope rests
in a space point of the rotating frame (this can make sense in
general relativity, too).

We recognize that the latter is analogous to the famous Foucault experiment:
the gyroscope is the pendulum (the base point of the gyroscope is the fixed
point where the pendulum is hung, the three axes of the gyroscope are:
vertical, horizontal in the plane of swings and perpendicular to the plane of
swings), the rotating reference frame is the Earth; the pendulum keeps its
direction (the direction of its axes) and precesses with respect to the Earth.
Therefore we find it convenient to introduce the terminology {\it Foucault
precession} to describe this situation.

{\it Thomas rotation is a strictly relativistic phenomenon, while
Foucault precession is a phenomenon which appears already in the
non-relativistic setting}. Under what circumstances can we use
Foucault precession to evaluate the Thomas rotation angle? In \cite{rindler}
the authors restrict themselves to circular orbits in axially symmetric
stationary spacetimes, and rotating frames with respect to which the metric is
also stationary. The authors of \cite{herrera} do not take sufficient care in
using the concept of \cite{rindler} and arrive at three different Thomas
rotation angles using three different rotating frames in special relativity
(we will see later what goes wrong there). Here we highlight the main points of
the concept of \cite{rindler}:

$a$; \ The world line $r$ of the base-point of the gyroscope is given, and
for some proper time instant $s$ there is a `natural' isometric
identification given between the tangent spaces at $r(0)$ and $r(s)$, such
that under this identification $\dot r(s)=\dot r(0)$,

$b$; \ There is a reference frame $\U$ co-moving with the gyroscope, i.e.
$r$ is an integral curve of $\U$ (a space-point in $\U$-space),

$c$; \ The Foucault-precession of the gyroscope in $\U$-space makes sense
(in some well-defined mathematical sense),

$d$; \ The Foucault-precession of the gyroscope is the negative of the
angular velocity of $\U$,

$e$; \ The correspondence of the tangent spaces at $r(0)$ and $r(s)$ established
by the Lie transport corresponding to $\U$ is equal to the `natural' identification
previously given.

In \cite{rindler}, in condition $a$ the world line $r$ is circular in an
axially symmetric stationary spacetime (Minkowski, Schwarzschild, Kerr,
G\"odel), and the `natural' identification is given by the Lie transport
corresponding to the `natural' coordinatization of the spacetime model. In
condition $b$ the rotating frame is selected so that it also defines a
stationary coordinatization, and $c,d,e$ are then (heuristic) consequences
of this choice.

In Sections \ref{fopre} and \ref{angvel} below we will give a mathematically
rigorous definition of Foucault precession, and give necessary and sufficient
conditions for $c,d$ above to hold.

In Section \ref{rotobs} we examine uniformly rotating reference frames in
Schwarzschild and Minkowski spacetimes, and find that the Foucault
precession makes sense for the conventional frame (as in \cite{rindler}), but
not for the `Trocheris-Takeno' and `modified Trocheris-Takeno' frames
(as in \cite{herrera}). In Section \ref{utolso} we show by examples that
condition $e$ above is not an automatic consequence of conditions $a,b,c,d$
even in Minkowski space, and therefore it needs to be taken into account
whenever applying the method of \cite{rindler}.

\section{Notions and notations}

We shall use a coordinate free formulation of relativity
(see e.g. \cite{sachs}), applying the following notions and notations:
A reference frame $\U$ is a four-velocity field in spacetime $M$. The flow
generated by $\U$ is denoted by $\RR\times M\to M, \quad (t,x)\mapsto R_t(x)$.
For fixed $t$,  $\D R_t(x)$ is the derivative
of $x\mapsto R_t(x)$ at $x$. For fixed $x$, $t\mapsto R_t(x)$ is a world line
function describing a maximal integral curve of $\U$.

A maximal integral curve of $\U$ represents a `reference particle'; the
set of reference particles constitutes the physical space of the reference
frame; so, a maximal integral curve is considered to be a {\it space point}
of $\U$.

Let the world line function $r$ describe a $\U$-space point,
i.e. $r(s)=R_s(x_0)$ for some world point $x_0$; considering $x_0$ as fixed,
we shall omit it from the following notations. Then

\be\label{ls}
\L(s):=\D R_s(x_0)
\end{equation}
is a linear bijection from the tangent space at $r(0)$ onto the tangent space
at $r(s)$ and

\be\label{rlie}
\dot r(s)= \L(s)\dot r(0).
\end{equation}

$\E_{\dot r(s)}$ will denote the linear subspace of the tangent space
at $r(s)$, orthogonal to $\dot r(s)$ and
\be
\Pp(s):=\1 + \dot r(s)\otimes\dot r(s)
\end{equation}
is the orthogonal projection onto $\E_{\dot r(s)}$.

\be\label{as}
\A(s):=\Pp(s)\L(s)\Pp(0)=\Pp(s)\L(s)
\end{equation}
is a linear bijection from $\E_{\dot r(0)}$ onto $\E_{\dot r(s)}$
(the second equality follows from \eqref{rlie}),

\be\label{as1}
\A(s)^{-1}:=\Pp(0)\L(s)^{-1}\Pp(s) = \Pp(0)\L(s)^{-1}
\end{equation}
is a linear bijection from $\E_{\dot r(s)}$ onto $\E_{\dot r(0)}$
(the second equality follows from \eqref{rlie}) and

\be\label{aao}
\A(s)^{-1}\A(s)=\Pp(0),\qquad \A(s)\A(s)^{-1}=\Pp(s).
\end{equation}

A vector field $\vv$ along $r$ is Lie transported according to $\U$ if
\be
\vv(s)=\L(s)\vv(0).
\end{equation}

A vector field $\vv$ along $r$ is space like Lie transported according to $\U$
if

\be
\vv(s)=\A(s)\vv(0).
\end{equation}

\section{Gyroscopes}\label{gyroscopes}

A gyroscopic vector is a Fermi-Walker transported space like vector field $\z$
along a world line function $r$  (which is called the base-point of the
gyroscopic vector). A gyroscope is a collection of three orthogonal gyroscopic
vectors $\z_i$ $(i=1,2,3)$ having the same base-point $r$.

\subsection{Thomas rotation}

Let us consider a gyroscopic vector $\z$ along $r$. As described in the
Introduction, Thomas rotation at some proper time instant $s$ is meaningful
only if

1) \ the tangent spaces at $r(0)$ and $r(s)$ are identified,

2) \ $\dot r(0)=\dot r(s)$ according to the identification in question.

Then, taking a gyroscope, the linear map defined by $\z_i(0)\mapsto\z_i(s)$
$(i=1,2,3)$ is the {\it Thomas rotation on the
world line $r$ between its proper time values $0$ and $s$}.

\begin{figure}

\center{\includegraphics[scale=0.5]{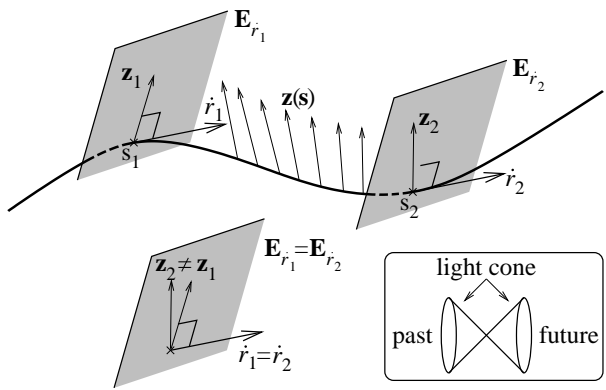}}

\caption{\label{f:t_r}Thomas rotation in SR. At two different proper time
values $s_1$ and $s_2$ the 4-velocities $\dot r_1 =\dot r(s_1)$
and $\dot r_2 =\dot r(s_2)$ are equal, so $\E_{\dot r_1} =\E_{\dot r_2}$,
but the initial and final gyroscopic vectors $\z_1 =\z(s_1)
\in\E_{\dot r_1}$ and $\z_2 =\z(s_2) \in\E_{\dot r_2}$ are different.}
\end{figure}

Note that condition 2) is important even in special relativity where
condition 1) is trivially satisfied.

In the cases examined by \cite{rindler}, condition 1) follows from
the stationarity and condition 2) follows from the axial symmetry.

\subsection{Foucault precession}\label{fopre}

Let $\U$ be a reference frame and $\z$ a gyroscopic vector along a
world line function $r$ which describes an integral curve of $\U$, i.e.
$\dot r(s)=\U(r(s))$. In other words, the base point of the gyroscope rests
in a space point of the reference frame.

The reference frame observes the gyroscopic vector as a time dependent
vector $\h(s)$ in  the $\U$-space point in question;
time, of course, means proper time of $r$.

Recall that $\U$-space is endowed with a smooth structure (\cite{matolcsi1}),
$\h(s)$ is a vector in the tangent space at the $\U$-space point; according to
the definition of that smooth structure, the tangent space in question can be
represented by the local rest frame at $r(0)$, and then  $\h(s)$ will be
represented by $\h_0(s):=\A(s)^{-1}\z(s)$.

Then we infer that
\be\label{hpont}
\dot\h_0 = (\A^{-1})^{\cdot} \z +
\A^{-1}\dot\z = -(\A^{-1}\dot\A)\h_0
\end{equation}
where, of course, the dot on the right hand side denotes $\nabla_{\dot r}$;
the second term in the middle expression is zero because $\dot\z$
is parallel to $\dot r$.

{\it The Foucault precession}, i.e. the angular velocity of a gyroscope with
respect to the
reference frame, {\it is meaningful if and only if $\Om_0(s):=-\A^{-1}(s)\dot\A(s)$
is an antisymmetric map in $\E_{\dot r(0)}$}, equivalently, {\it if and only if
$\A(s)$ is an isometric map from $\E_{\dot r(0)}$ to $\E_{\dot r(s)}$}
for all $s$. If so, then it is natural to define the Foucault precession at
$s$ as the antisymmetric map
\be\label{foucpre}
\Om(s):=\A(s)\Om_0(s)\A^{-1}(s) =-\dot\A(s)\A^{-1}(s).
\end{equation}

\begin{figure}
\center{\includegraphics[scale=0.5]{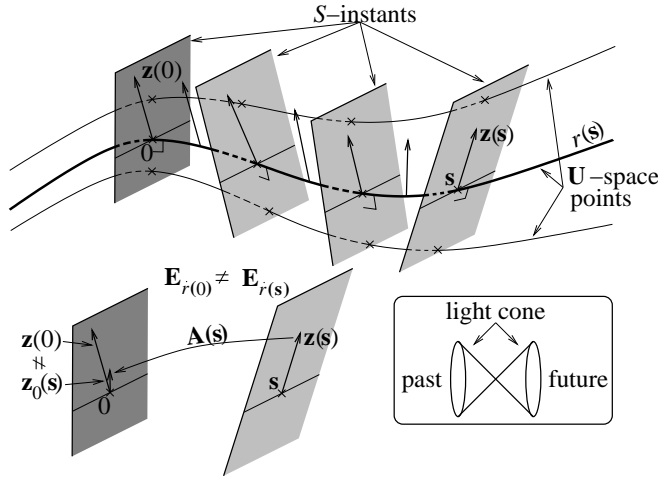}}
\caption{\label{f:f_p}Foucault precession. A
reference frame $\U$ perceives a precession of a gyroscopic vector
$\z$ whose base-point rests at the space point $r$ of the reference frame.}
\end{figure}

\subsection{Angular velocity of reference frames}\label{angvel}

According to the usual definition, {\it the angular velocity} (which is also
often referred to as vorticity) {\it of a reference frame} $\U$   is

\begin{equation}\label{e:ang}
\Om_\U:= -\frac1{2}(1+\U\otimes\U)(\D\land\U)(1+\U\otimes\U).
\end{equation}

Thus, the angular velocity of the reference frame at $r(s)$ equals
\be
-\frac1{2}\Pp(s)(\D\land\U)(r(s))\Pp(s).
\end{equation}

The reference frame is defined to be rigid along $r$ (see \cite{sachs}, p. 56),
if $\Pp(s)(\D\U)(r(s))\Pp(s)$ is antisymmetric. Then its
angular velocity at $r(s)$ is
\be\label{e:angvel}
\Pp(s)(\D\U)(r(s))\Pp(s)=\Pp(s)\dot\L(s)\Pp(s).
\end{equation}

We can now deduce from formulae  \eqref{foucpre} and  \eqref{e:angvel} that {\it the Foucault precession is meaningful} at a space
point of a reference frame $\U$ {\it if and only if $\U$ is rigid at that
space point, and then the angular velocity of the Foucault precession is
the negative of the angular velocity of the reference frame}.
Indeed, since the zero proper time point can be chosen arbitrarily, we
have to show only that $\dot\A(0)\A^{-1}(0)=\Pp(0)\dot\L(0)\Pp(0)$ which follows
from \eqref{as}, \eqref{as1} and from the facts that $\L(0)$ is the identity
and $\Pp(0)\dot\Pp(0)\Pp(0)=0$.

Subsections \ref{fopre} and \ref{angvel} clarify the theoretical background of
conditions $c, d$ in the Introduction. The necessary and sufficient condition
for $c$ and $d$ to hold is that the reference frame co-moving with the
gyroscope be {\it rigid in the base-point of the gyroscope}.

\section{Rotating reference frames} \label{rotobs}

In this section we examine uniformly rotating reference frames in  Minkowski
and Schwarzschild  spacetimes. An abstract definition is given in both cases,
but further calculations are only carried out in Minkowski spacetime to make
comparison with \cite{herrera} possible.

Here we use the notations of the special relativistic spacetime model expounded in
\cite{matolcsi1}, \cite{matolcsi2} (which is in accordance with \cite{sachs}).
Minkowski spacetime $M$ is
an affine space over the vector space $\M$ and the spacetime
distances form an oriented one dimensional vector space $\I$, so
the metric tensor is of the form $\M\times\M\to\I\otimes\I$,
$(\x,\y)\mapsto\x\cdot\y$.

The Levi-Civita connection is the usual differentiation, denoted by
the symbol $\D$.

In Minkowski spacetime, in this abstract formalism one can easily introduce a
general notion (a whole family) of uniformly rotating reference frames,
and the usually considered rotating coordinate transformations  appear as
special cases of this general form (with appropriate synchronization).

It is remarkable that finding the `canonical' coordinate system describing a
rotating reference frame has been a minor but long-standing problem in special
relativity theory; that is why the `Trocheris-Takeno' (\cite{trocheris},
\cite{takeno}) and the `modified
Trocheris-Takeno' transformations (\cite{herrera2}) were introduced, after
theorists were not satisfied with the conventional rotating coordinates.

In coordinate transformations, however, a reference frame and a synchronization
are intermingled and this can lead to confusion: e.g. the Trocheris-Takeno
transformation and the modified
Trocheris-Takeno transformation concern the same reference
frame with different synchronization.

It is also worth mentioning here that rotating reference frames seem to give
rise to numerous misunderstandings and alleged paradoxes in special relativity
(see e.g. the Ehrenfest paradox \cite{ehren} with a number of refutations
from different points of view \cite{ehren1, ehren2, ehren3, ehren4},
or Selleri's recent paradox \cite{selleri}  with refutations in
\cite{matrotating}, \cite{rodrigues}.) We feel that all these
misunderstandings could be avoided if the standard language of relativity
(i.e. coordinate systems and coordinate transformations) were replaced by
a systematic use the abstract notions of reference frames and synchronizations.

Heuristically a (uniformly) rotating reference frame is characterized by the
property that its space points are rotating around an inertial centre
which is the world line $o+\uu\I$, described by a specific world point $o$,
and a four-velocity $\uu$. The rotation around the
centre, i.e. in the  spacelike hyperplane $\Eu$, is characterized by
the angular velocity of the rotation, an antisymmetric linear map
$0\neq\Om:\Eu\to\frac{\Eu}{\I}$ which is
conveniently extended to the whole $\M$ in such a way that $\Om\uu=0$.
Then at an arbitrary point $x\in M$ the velocity of the
rotation relative to the centre is proportional to
$\Om(x-o)$, so $\U(x)$ is the linear combination of $\uu$
and $\Om (x-o)$.
Thus, we accept that given positive real valued smooth functions
$a$,$b:\RR^+\to \RR^+$ such that
\be\label{e:a2}
a(x)^2 -b(x)^2|\Om(x-o)|^2=1,
\end{equation}
a corresponding rotating reference frame is defined as
\begin{equation}\label{e:uro}
\U(x)=a(x) \uu +b(x)\Om(x-o).
\end{equation}
The normalization condition \eqref{e:a2} ensures that $\U$ does  indeed
map to the set of four-velocities.

Note the following special cases

1. $a(x)=b(x)=\dfrac1{\sqrt{1-|\Om(x-o)|^2}}$
which is the conventional rotating reference frame
(\cite{moller}, \cite{matolcsi2}),

2. $a(x)=\cosh|\Om(x-o)|$,
$b(x)=\dfrac{\sinh|\Om(x-o)|}{|\Om(x-o)|}$
(the Trocheris-Takeno reference frame \cite{trocheris}, \cite{takeno}),

3. $a(x) = \sqrt{1+|\Om(x-o)|^2}$, $b=1$ (\cite{matolcsi2}),

4. $a=\mathrm{const}>1$, $b(x)=
\dfrac{\sqrt{a^2-1}}{|\Om(x-o)|}$.

It is a simple fact that all the $\U$-space points are circular world lines:
the one passing through the world point $x$ is given by the function
\be
s\mapsto  o + s a(x)\uu + e^{s b(x)\Om}(x-o). \label{e:ulinx}
\end{equation}

Now let us consider Schwarzschild spacetime which can be given by the previous
objects in such a way that the spacetime metric is
\be
\mathbf g(x):=\1 + h(x)\uu\otimes\uu + \frac{h(x)}{1-h(x)}\mathbf n(x)\otimes
\mathbf n(x)
\end{equation}
where
\be
h(x):=\frac{2m}{r(x)}, \quad \mathbf n(x):=\frac{(\1+\uu\otimes\uu)(x-o)}{r(x)},
\end{equation}
\be
r(x)=|(\1+\uu\otimes\uu)(x-o)|.
\end{equation}

Now, if $(1+h(x))a(x)^2 -b(x)^2|\Om(x-o)|^2=1$ then \eqref{e:uro} defines a
rotating reference frames for world points $x$ satisfying $r(x)>2m$. For further elaboration and an interesting application of
this formalism and rotating observers in Schwarzschild spacetime see \cite{gps}, where the time-rate between satellite clocks and Earth-based clocks is calculated.

\subsection{Foucault precession in rotating reference frames}
\label{fukorot}

Now we investigate whether Foucault precession in the space of
the rotating reference frames defined above in Minkowski spacetime is
meaningful. In view of Section \ref{gyroscopes} this is equivalent to the reference frames being rigid.

We restrict our attention to the case when the coefficients
in the linear combination depend only on $|\Om(x-o)|^2$ (and not on $x$),
i.e. there are positive real valued smooth functions
$\al,\bb:\RR^+\to \RR^+$ such that
\be
a(x)=\al(|\Om(x-o)|^2),\qquad b(x)=\bb(|\Om(x-o)|^2).
\end{equation}
For the sake of brevity, we introduce the notation
\be
k(x):=|\Om(x-o)|^2
\end{equation}
and we shall use the following formulae:
\be\label{e:Dk}
\frac{dk(x)}{dx}=\frac{d|\Om(x-o)|^2}{dx}=-2\Om^2(x-o),
\end{equation}

\begin{align}\label{e:Dhab}
\frac{d\al(k(x))}{dx}&=-2\al'(k(x))\Om^2(x-o), \\
\frac{d\bb(k(x))}{dx}&=-2\bb'(k(x))\Om^2(x-o),
\end{align}
where the prime denotes differentiation with respect to the real variable
of the functions. Moreover, we infer from \eqref{e:a2} that
\begin{equation}\label{aaderiv}
2\al(k)\al'(k)-2\bb(k)\bb'(k)k=\bb^2(k).
\end{equation}

Then we find that
\begin{equation}
\D\U(x)=-2\bigl(\al'(k(x))\uu+\bb'(k(x))\Om(x-o)\bigr)
\otimes \Om^2(x-o)+ \bb(k(x))\Om,
\end{equation}

A simple calculation shows that
$(\1+\U(x)\otimes\U(x))\D\U(x)(\1+\U(x)\otimes\U(x))$ is
antisymmetric if and only if
\begin{equation}\label{alfbet}
 2\al'=\al\bb^2 \qquad 2\bb'=\bb^3.
\end{equation}

We can solve the second equation for $\bb$,  and then
taking into account (\ref{e:a2}), we find that
there is a positive constant $a$ such that
\begin{equation}\label{joforgo}
 \al(k(x))=\frac1{\sqrt{1-a^2|\Om(x-o)|^2}}, \qquad
\bb(k(x))=\frac{a}{\sqrt{1-a^2|\Om(x-o)|^2}}. \end{equation}

Therefore, we conclude that the Foucault precession is meaningful
for the conventional rotating frame ($a=1$), but it is {\it not meaningful}
for the rotating reference frames 2, 3 and 4 listed in Section \ref{rotobs}.
Hence, {\it the principle of \cite{rindler} can be applied to the conventional
rotating frame} (where condition $e$ is also satisfied, as is easy to check),
{\it but  not to the `Trocheris-Takeno' or `modified Trocheris-Takeno' frames}.

\section{Thomas rotation versus Foucault precession}\label{utolso}

In this section we show that the angular velocity of a single world line
cannot canonically be defined. This shows that the Foucault precession of a
gyroscope depends on the choice of the co-moving reference frame. Accordingly,
the knowledge of instantaneous Foucault precessions can only be used to
determine the Thomas rotation angle if condition $e$ of the Introduction is
satisfied. We will show by examples that this condition is not a consequence of
$a,b,c,d$ even in Minkowski space, and therefore it needs to be taken care of
separately whenever applying the principle of \cite{rindler}.

\subsection{A special family of reference frames}\label{noang}

Let $r$ be an arbitrary smooth world line function in Minkowski spacetime.
For $x$ in a neighbourhood of the range of $r$, there is a unique
proper time value $s(x)$ of $r$ such that $x-r(s(x))$ is orthogonal
to $\dot r(s(x))$; it is determined by the implicit relation
$(x-r(s))\cdot\dot r(s)=0$.

The function $x\mapsto s(x)$  satisfies
\begin{equation}\label{implido}
\frac{ds(x)}{dx}= - \frac{\dot
r(s(x))}{1+(x-r(s(x)))\cdot\ddot r(s(x))},
\end{equation}

Note that
\be\label{rimplido}
\frac{ds(x)}{dx}\Big|_{x=r(s)}=-\dot r(s).
\end{equation}

Let $s\mapsto\Hh(s)$ be a smooth map such that
$\Hh(s):\M\to\M$ is Lorentz transformation for which $\Hh(s)\dot r(0)=
\dot r(s)$ holds. ($\Hh(s)$ can be for example the Lorentz boost from
$\dot r(0)$ to $\dot r(s)$ or the Fermi-Walker transport along $r$ from $0$ to
$s$.) Given such a one-parameter family $\Hh(s)$, and any antisymmetric
linear map $\Gamma:\M\to\frac{\M}{\I}$ for which $\Gamma\cdot\dot r(0)=0$,
the associated family $\Hh_\Gamma (s):=\Hh(s)e^{s\Gamma}$ is another
such family, so we have some freedom when choosing $\Hh(s)$.

Then putting
\be
\V(x):= \dot r(s(x)) +
\dot\Hh(s(x))\Hh(s(x))^{-1}\bigl(x-r(s(x))\bigr),
\end{equation}
we define the reference frame

\be\label{ellenmf}
\U(x):=\frac{\V(x)}{|\V(x)|}
\end{equation}
where, of course, $|\V|=\sqrt{-\V\cdot\V}$. Evidently, $\dot r(s)=\U(r(s))$,
so the world line described by $r$ is the space point of different reference
frames given by different $\Hh$-s.

Then
\be
\D\U=\frac{\D\V}{|\V|} +\frac{\V\otimes(\D\V)\V}{|\V|^3}.
\end{equation}
Since $\V(r(s))=\dot r(s)$, $|\V(r(s)|=1$,
$\D\V(r(s))= -\ddot r(s)\otimes\dot r(s) +\dot \Hh(s) \Hh(s)^{-1}
(1 +\dot r(s)\otimes\dot r(s))$ and $(\D\V(r(s))\V(r(s)=\ddot r(s)$,
we have that
\be
\Pp(s)(\D\U)(r(s))\Pp(s) = \Pp(s)\dot \Hh(s)\Hh(s)^{-1}\Pp(s).
\end{equation}
Since $\dot \Hh(s)\Hh(s)^{-1}$ is antisymmetric, the right hand side
is antisymmetric as well, thus the reference frame is rigid at $r$ and its
angular velocity (see \eqref{e:angvel}) at $r(s)$ is
\be\label{angobs}
\Pp(s)\dot \Hh(s)\Hh(s)^{-1}\Pp(s).
\end{equation}
If we take $\Hh_\Gamma (s):=\Hh(s)e^{s\Gamma}$, then the
angular velocity of the reference frame at $r(s)$ is
\be\label{gamma}
 \Pp(s)\dot\Hh(s)\Hh(s)^{-1}\Pp(s) + \Hh(s)\Gamma\Hh(s)^{-1}.
\end{equation}

Thus, according to our result in Subsection \ref{angvel}, for any
choice of $\Hh(s)$, the Foucault precession in the space point, given by $r$,
of the reference frame \eqref{ellenmf} is meaningful.

We call attention to the following fact. According to \eqref{gamma},
{\it the same world line can be a space point of
different reference frames with different angular velocities; the properties of
a single world line are in no relation with the angular
velocity of a reference frame having the world line as a space point;
the angular velocity of a single world line cannot be defined}.

Note that, in particular,

-- \ if $\Hh(s)$ is the Lorentz boost from $\dot r(0)$ to
$\dot r(s)$, then $\dot\Hh(s)\Hh(s)^{-1}=\\
(\dot r(s)+\dot r(0))\land\ddot r(s)$ and the angular velocity of the
refrence frame at $r(s)$ equals \\ $\bigl(\dot r(0) + \dot r(s)(\dot r(s)
\cdot\dot r(0))\bigr)\land\ddot r(s)$,

-- \ if $\Hh(s)$ is the Fermi-Walker transport along $r$
from $0$ to $s$, then $\Hh(s)^{-1}\dot\Hh(s)$ $=\dot r(s)\land\ddot r(s)$
and the angular velocity of the reference frame at $r(s)$ is zero.

One can be tempted to say that the circular world line \eqref{e:ulinx}
has angular velocity  $\bb(x)\Om$ but in view of the discussion above
this is not right: the circular world line can be
the space point of several reference frame having different angular velocities.

\subsection{Thomas rotation versus Foucault precession}\label{contra}

If the Foucault precession of a gyroscope with respect to a co-moving reference frame
is meaningful, then the time-integral of the Foucault precession results in a
finite angle. This angle can then be compared to the Thomas rotation angle.
We have seen in the Introduction that in Special Relativity the instantaneous
Thomas precession depends on the inertial frame who observes the gyroscope, but
its time integral always gives the Thomas rotation angle when the gyroscope returns
to its initial local rest frame. One might be tempted to think that
the situation is similar with Foucault precession:  although it depends on the
choice of the co-moving frame, the calculated rotation angle should always give
the Thomas rotation.  However, this is not true in such generality and some
care must be taken here.

In fact, after an arbitrary time period $\s$, the Foucault precession results
in an angle whose cosine is $\frac{\h_0(0)\cdot\h_0(s)}{|\h(0)|^2}=
\frac{\z(0)\cdot\A(s)^{-1}\z(s)}{|\z(0)|^2}$,
$\h_0$ being the solution of the differential equation \eqref{hpont}.
Thomas rotation, on the other hand, is only defined at such an $s_T$ for
which some previously given `natural' identification $\N$ is given between the
tangent spaces at $r(0)$ and $r(s_T)$. In that case, the cosine of Thomas
rotation is given by
$\frac{\z(0)\cdot\N^{-1}\z(s_T)}{|\z(0)|^2}$. Therefore, the necessary and
sufficient condition for the two angles to be the same is that
$\A(s_T)^{-1}\z=\N^{-1}\z$ for all space-like vectors $\z$, i.e.
$\A(s_T)=\N\Pp(0)$. This is what we called condition $e$ in the Introduction.
We now show by an easy example that {\it this condition is not a direct consequence of the
other conditions} $a,b,c,d$, even in Minkowski spacetime where the identification
$\N$ is taken to be the identity.

Indeed, let us consider the reference frame \eqref{ellenmf}.
It is easy to see that the function $s\mapsto\rho_\q(s):=r(s) +\Hh(s)\q$
for $\q\in\E_{\dot r(0)}$ satisfies $\dot\rho_\q(s)=\V(\rho_\q(s))$.
Thus, though $\rho_\q$ is not a world line function, its range is a world line,
a space point of $\U$. The proper time value $t$ corresponding to the
world line described by $\rho_\q$ defines a proper time value $s(t,\q)$ of
$r$ via the differential
equation $\frac{ds}{dt}=\frac1{|\dot r(s) + \dot\Hh(s)\q|}$.

As a consequence, we have for the flow generated by $\U$:
\be
R_t(r(0)+\q)= r(s(t,\q)) + \Hh(s(t,\q)\q \qquad
(\q\in\E_{\dot r(0)}).
\end{equation}

For getting $\A(s)$ (see \eqref{ls} and \eqref{as}), we have
to differentiate the flow in the plane $r(0)+\E_{\dot r(0)}$, i.e. to
differentiate the above expression with respect to $\q$ and then to put
$\q=0$. As a result we get

\be\label{has}
\A(s)=\Hh(s)\Pp(0) =\Pp(s)\Hh(s)\Pp(0)=\Pp(s)\Hh(s).
\end{equation}

Similarly, we have $\A_\Gamma(s)=\Hh(s)e^{s\Gamma}\Pp(0)$ for the reference frame
constructed by $\Hh_{\Gamma}(s):=\Hh(s)e^{s\Gamma}$. Clearly, $\A_\Gamma(s)$ depends on the choice of $\Gamma$ and need not equal the identity operator. Therefore, condition $e$ is not a consequence of $a,b,c,d$.

\section{Conclusion}

We conclude that the principle of \cite{rindler} to calculate the Thomas
rotation angle of a gyroscopic vector is applicable whenever conditions
$a,b,c,d,e$ are satisfied. Conditions $c$ and $d$ are equivalent to the fact
that the co-moving reference frame is {\it rigid at the base point of
the gyroscope}, while condition $e$ has
to be checked separately. All conditions are satisfied in the cases considered
in \cite{rindler} due to the symmetries of the spacetime models, the circularity
of the world line, and the choice of the co-moving rotating frame. In
\cite{herrera} the `Trocheris-Takeno' and the 'modified Trocheris-Takeno'
frames fail to be rigid, and hence lead to an incorrect Thomas rotation angle.
It is also clear from our discussion that the method of \cite{rindler} is,
in principle, {\it not restricted} to stationary spacetimes or reference frames
corresponding to Killing vector fields. For instance, the Thomas rotation of
a gyroscope moving along an elliptic world line in Schwarzschild spacetime
could also be calculated if a rigid co-moving frame with property $e$ is
given (which, however, does not seem easy to define).

\end{document}